\documentclass[conference]{IEEEtran}
\IEEEoverridecommandlockouts
\usepackage{cite}
\usepackage{amsmath,amssymb,amsfonts}
\usepackage{graphicx}
\usepackage{xcolor, soul}
\usepackage{multirow}
\usepackage{setspace}
\usepackage{lineno,hyperref}
\usepackage{algorithmic,algorithm} 
\usepackage{float}
\usepackage{graphicx}
\newfloat{algorithm}{t}{lop}
\usepackage[utf8x]{inputenc}
\usepackage[tight,footnotesize]{subfigure}
\usepackage{stfloats}
\usepackage{url}
\usepackage{amssymb}
\usepackage{rotating}
\usepackage{tabularx}
\usepackage{adjustbox}
\usepackage{textcomp}
\usepackage{subfigure}
\usepackage{calrsfs}
\newtheorem{definition}{Definition}
\def\BibTeX{{\rm B\kern-.05em{\sc i\kern-.025em b}\kern-.08em
    T\kern-.1667em\lower.7ex\hbox{E}\kern-.125emX}}
\begin{document}

\title{Data Behind the Walls – An Advanced Architecture for Data Privacy Management}

\author{\IEEEauthorblockN{Amen Faridoon*}
\IEEEauthorblockA{\textit{School of Computer Science} \\
\textit{University College Dublin}\\
Dublin, Ireland \\
amen.faridoon@ucdconnect.ie}
\and
\IEEEauthorblockN{M. Tahar Kechadi}
\IEEEauthorblockA{\textit{School of Computer Science} \\
\textit{University College Dublin}\\
Dublin, Ireland \\
tahar.kechadi@ucd.ie}
}
\maketitle
\begin{abstract}
In today's highly connected society, we are  constantly asked to provide  personal information to
  retailers, voter surveys, medical professionals, and other data collection efforts.  The collected
  data is stored in large data warehouses. Organisations and statistical agencies share and use this
  data  to facilitate  research in  public health,  economics, sociology,  etc.  However,  this data
  contains sensitive  information about individuals, which  can result in identit theft, financial
  loss,  stress and  depression,  embarrassment, abuse,  etc. Therefore,  one  must ensure  rigorous
  management of individuals' privacy.  We propose,  an advanced data privacy management architecture
  composed  of  three  layers.   The  data   management  layer  consists  of  de-identification  and
  anonymisation, the access management  layer for re-enforcing data access based  on the concepts of
  Role-Based Access Control and the Chinese Wall Security Policy, and the roles layer for regulating
  different users. The proposed system architecture is validated on healthcare datasets.
\end{abstract}


\section{INTRODUCTION}
\label{sec:Int}
New technologies, such  as artificial intelligence, cloud computing, sensors  and wireless networks,
are constantly  integrated into our daily  lives at a  rapid pace. We  live in a digital  word, also
called the fourth industrial revolution.  This digital world produces a considerable amount of data,
which  is highly  valuable  and  can be  seen  as  the new  gold  or the  new  oil  of this  century\footnote{https://www.cutimes.com/2020/07/17/big-data-the-new-oil-fields/?slreturn=20220716063653}. 
This  data enables businesses or  organisations to enhance their  productions and
profits \cite{fisher2009data}.  Many data organisations, such as Facebook, Amazon, Google, ..., have
emerged as dominant in  this field.  They invested a lot of resources  to create, acquire, maintain,
and   store  the   data  to   make   constructive  and   timely  decisions   and  improve   customer
services. Therefore, with the rise of the data economy, companies find enormous value in collecting,
sharing, and using data. 

Data  breaches and  data theft  affect  both individuals  and organisations.   Most businesses  have
elaborated     security     mechanisms     to     protect     individual     private     information
\cite{greenaway2015company}.   However,  due  to   vulnerabilities  in  software,  phishing,  stolen
credentials, and  malicious attacks, cybercriminals  can infiltrate  and steal personal  data, which
results in  identity theft,  financial loss, stress  and depression,  discrimination, embarrassment,
abuse,  etc.  \cite{chabot2015ontology}  \cite{van2006challenge}.  Many  businesses  and  government
agencies have suffered a  massive data breach leading to layoffs, lawsuits, and  frauds. In 2016 IBM
reported that around $60\%$ of company data breaches were made by its employees, out of which $75\%$
were deliberate\footnote{https://www.ciab.com/resources/ibm-60-percent-attacks-carried-insiders/}.

An insider threat is  defined as any malicious activity executed by users  with legitimate access to
the organisation's network,  applications, or data stores \cite{huth2013guest}.  One can distinguish
three types of insider threats: 1) malicious insider who deliberately looks to steal information, 2)
careless  insider who  does  not follow  proper  IT procedures  and  accidentally exposes  sensitive
corporate data  at risk and 3)  compromised insider who  has compromised credentials by  an external
attacker. Unlike threats from  outside the organisation, insider threats are  much more difficult to
detect and prevent. Insider  threats' behaviour is difficult to identify  as suspicious because they
have legitimate  credentials and  access to the  systems and  data. Once inside  the network,  it is
difficult  to pick  apart the  malicious  behaviour from  a  large volume  of legitimate  day-to-day
activities.  The data  requires robust  privacy  management systems  to prevent  identity theft  and
unauthorised access from external and malicious insider attackers.

Therefore, the main challenge of this study is to design a data privacy management architecture that
can  prevent breaches  of  an individual's  identity  and  keep the  data  secure from  unauthorised
access. The  main objectives  of our  data privacy  management architecture  are twofold:  1) create
security walls between subjects and objects for access control and formulate authorisation rules for
legitimate users, and  2) secure the identity  and sensitive information of an  individual from both
the third parties and insider threats. 

\section{Related Work}
\label{sec:RW}
Numerous data privacy management models have been proposed and implemented by addressing the problem of unauthorized access control whereas the most common are; Mandatory Access Control (MAC), Discretionary Access Control (DAC), Attribute Based Access Control (ABAC) etc. However, these models have limitations. MAC is a very inflexible security model where each user is assigned a clearance, and each object has a classification and compartment stored in its security label. This model is mostly suitable for structured environments where we have classified data. DAC is less central system administration model because the control is not dictated by a company policy but rather by the owner or creator of the information. Whereas, the drawbacks of ABAC limits its implementation such as; time-consuming initial setup in term of identifying right resources, objects and subjects attributes, policies etc, cannot factor everything in environments like big organizations, and it also not ideal for small businesses. \\
According to the organizational structure and needs the most considerable models is Role Based Access Control (RBAC). Most of the studies \cite{zhou2013achieving} \cite{shahid2021big} \cite{nguyen2017privacy} \cite{tran2019security} took an advantage of RBAC security model and their extended versions in different environments (cloud, distributed data storage systems etc.) to restrict the anuthorized access. Whereas, in the proposed systems the administration of permission groups having
the  main  responsibility for  managing  the  users, their  roles  and  permissions.  Therefore,  by exploiting their authority, the access permissions of the roles can easily be changed. \\
Moreover, researchers also adopt this model in healthcare domain. Authors in \cite{blobel2004authorisation} identified the need  for roles and their  acting procedures in an  EHR system. They highlighted that the processes for  defining the roles in an EHR system play  a vital part in access privileges management. Others proposed structured roles and grant access only to entities located in the authorisation table \cite{france2004security}.  The privacy and  security maintenance committee is the only authority that can act on  the authorisation table. However, structured roles are not  enough, and the dynamic and structural  aspects of  roles are  necessary \cite{van2003data}. For  instance, one  should only  provide data access to physicians  if they are engaged in  their treatment. Therefore, one can  deduce roles have more contextual variables (e.g.,  place, time, association, etc.). Hence, the  first model, known as contextual role-based access control authorisation, was proposed  in \cite{motta2003contextual}. The purpose of this model is to improve privacy  and confidentiality of patients' records, as they were not robust enough. The model considers a  hierarchy of roles with  authorisation inheritance and a data  model which covers the different  data types (e.g., prescriptions,  demographics). The model also  includes a technique for  authorisation conflict  management. However, most  of  these models consider access to healthcare practitioners only, they do not have explicit policies on the complete dataset. More  data professionals should be considered and be able to access the data with predefined usage and ethics.\\
The applications of Chinese Wall Security Policy (CWSP) in different environments have attracted significant research interests in the last few years. Studies deployed the  CWSP model on the cloud at the IaaS level to control the information flow between the conflicting parties \cite{xie2013information}\cite{wu2010information}. Moreover, various scenarios have been modelled with CWSP, mainly when conflict-of-interests are involved. For instance, studies \cite{fehis2016new}\cite{fehis2015new} focused on protecting subjects (companies) and objects (datasets) by
creating subject walls and object walls. Recently, a security  model for cryptographic key management in the cloud was proposed and offered security as a Service (CKMS$-$SecaaS’s) \cite{fehis2021secure}. \\
These studies, while  efficient, need further investigations because the spectrum of their applications is limited to distributed or cloud environments. Their main objective was to secure the datasets or assets of an organisation from third-party (conflicting organisation) attacks. Nevertheless, guaranteeing the privacy and security of personal and  sensitive datasets within an organisation from insider attacks is still challenging.
\section{Proposed Model Concepts}
\label{sec:PMC}
The  proposed model  consists  of  three modules:  the  roles (or  users)  module,  the data  access
management module,  and the  data management  module. We  combine the  concept of  role-based access
control  with  the  Chinese  wall  security  policy to  accomplish  data  security. The objective is  to manage data storage and usage so  that individuals’ data cannot be exposed  to malicious parties.   
\subsection{Roles, Subjects, and Objects}

The   Role-Based   Access   Control   model   was  proposed   in   1996   in   \cite{sandhu2000nist}.  
It is  considered one  of the  most accepted  access control  models. The  model exploits the  “need-to-know” principle, widely  adopted and utilised.  The whole model  assumes that privacy is preserved  as long as the data is  only accessed when necessary for the  right motive and
only  minimum  information  details  are  disclosed.

\begin{figure} [h]
  \centering
  \includegraphics[scale=0.5]{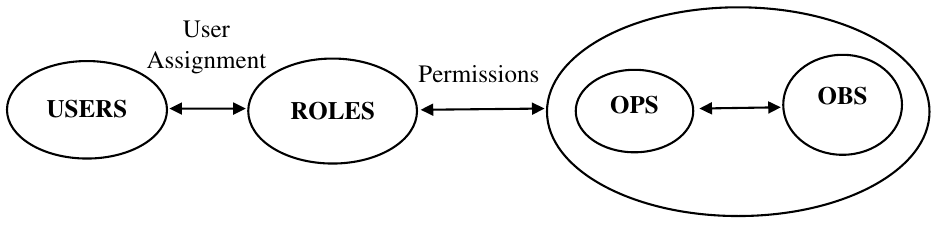}
  \caption {Core Role-Based Access Control Model}
  \label{fig:RBAC}
\end{figure}

The core components shown in Figure \ref{fig:RBAC} of the role-based access control model are Users or Subjects, Roles, Operations, and Objects. 

\begin{itemize}
  \item {\bf Users} are the skilled individuals and they are grouped into roles.
  \item {\bf Roles} represent a set of responsibilities defined by the administration of an                  organisation.
  \item {\bf Operations (OPS)} are functions performed by the subjects (users) on the objects.
  \item {\bf Objects (OBS)} are the resources that are managed and used by the subjects.
        A resource can be an application, database, file, etc.
\end{itemize}

\subsection{Access Management and Walls}
\label{ssec:AML}
\begin{figure} [h]
  \centering
  \includegraphics[scale=0.4]{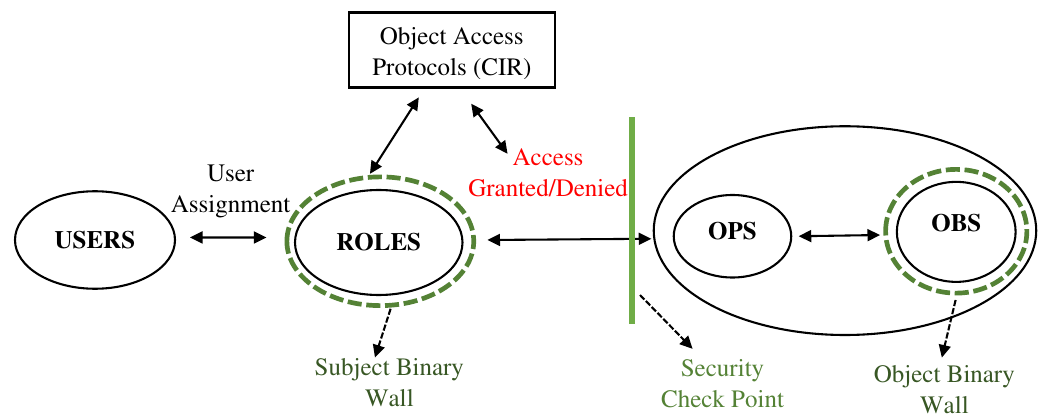}
  \caption {Data Access Control Based on CWSP}
  \label{fig:RBAC-CWSP}
\end{figure}

The proposed data access management, based on the Chinese Wall security policy (CWSP), is defined as
a dynamic  firewall mechanism shown  in Figure \ref{fig:RBAC-CWSP}.  We implemented the  concept of
CWSP  for data  access control  between competing/conflicting  roles based  on access  control rules
\cite{fehis2016new, fehis2015new, brewer1989chinese}. The walls defined in this model prevent access
between the roles and data warehouses inside the same organisation. The adoption of the CWSP for the
data access management layer is summarised below: 

\begin{itemize}
  \item Users/subjects  cannot access the object that is in conflict with the object  they already possess. Therefore, each subject  has a granted and  a denied objects,  where each subject in the granted set has their conflicting subjects in the denied set. The pairwise (Granted, Denied) of a subject is called the subject wall. It is not possible to find two competing objects inside the same wall.
  \item The  subjects can  perform read and write  operations via authorised access  to the objects. In this  case, the wall, called as “object wall”,  is created around the objects based on  the same rules as the subjects.
\end{itemize}

\subsection{Wall Placement}
Chinese wall security  policy is based on wall  placements to prevent access to  data platforms from
unauthorised subjects. Our model acts on the  same entities as previous ones: subjects, objects, and
access rights.  In this  study, we  have data  warehouses (DWs)  as objects  and a  set of  users of
different responsibilities  on DWs.  We achieve the  privacy of individuals’  records in  two steps:
de-identification and anonymisation of original data records. 

The  typical  data  platform  architectures  should deal  with  data  collection,  integration,  and
analytics. We  divide the collected data  stores into three main  layers, depending on the  level of
privacy and security they may need to handle. They are described in the following: 

\paragraph{Original Data Warehouse (ODW)}
The  collected  raw  data via  different  devices  and  from  various  sources are  stored  in  data
marts.  These  data  marts are  carefully  integrated  into  a  data warehouse  containing  original
data.  This data  warehouse maintains  the  quality, accuracy,  validity, timeliness,  completeness,
uniqueness, and  consistency of the  original data  records. We consider  this data warehouse  as an
object (ODW) that must be secured from unauthorised  access and the limited number of users who have
access to these original records. 

\paragraph{De-identified Data Warehouse (DDW)}
All  information  about   users’  identities  is  removed  to  prevent   identity  disclosure.  This
de-identified data warehouse consists of individual  aggregated and derived statistical data that is
computed from the original data. We consider  this de-identified data warehouse as a separate object
(DDW) that  needs to prevent the  identity of an entity  during analysis. But, with  the presence of
quasi-identifiers, this data needs further data protection. 

\paragraph{Anonymised Data Warehouse (ADW)}
After the  de-identification, we  further anonymise  the data so  that the  original data  cannot be
recovered from the  knowledge extracted during the  data analysis. We consider  this anonymised data
warehouse as an  object (ADW) that must be  secured from malicious third parties or  the linkage and
other attacks.  
\section{Model Formulation}
\label{sec:PMF}
In  the previous  section,  we  identified the  main  concepts of  the  model:  roles, objects,  and
walls. The  roles are the  subjects that own  the set of objects.  The objects
represent  datasets, data  records, or  data warehouses  (ODW, DDW and ADW). Finally,  the walls  implement  the conflicting  relation  on which  the CWSP  is
based. In the following, we give formal definitions of all these entities and relations.

The  overall system  can be  considered as  $(R,  O, CIR,  CIC, W, A)$,  where $R  = \{r_1,\cdots,  r_n\}$,
$O   =   \{o_1\cdots,   o_m\}$,   $CIR   =   \{cr_1\cdots,   cr_k\}$,  $CC   =   \{cc_1\cdots,   cc_p\}$, $W   =   \{w_1,\cdots,   w_l\}$,
$A = \{a_1,\cdots,  a_h\}$ sets of Roles (subjects), Objects,  Conflict-relations, conflict-classes, Walls, and access
rights, respectively. They are defined as follows:\\
\\
The proposed system has following characteristics:
\begin{enumerate}
  \item Some roles are in conflict:\\
  Let $ \{R_i\} $ and $ \{R_j\} $ are in conflict:
  if  u $ \in \{R_i\} $, then u $ \notin \{R_j\} $ and u cannot be in $ \{R_j\} $ at all. Therefore, we need a wall between $ \{R_i\} $ and  $ \{R_j\} $.
  \item Some roles are cooperative:\\
  Let $ \{R_x\} $ and $ \{R_y\} $ are cooperative:
  if  u $ \in \{R_x\} $, then u $ \notin \{R_y\} $ but u can switch to $ \{R_y\} $. So, there is no need to have a wall between $ \{R_x\} $ and  $ \{R_y\} $.
\end{enumerate}

Suppose, $D$ is a set of domains, $D = \{d_1, \cdots, d_n\}$, 
 $\Phi$ denotes Execution Denied, $\{O_{d_i}\}$ denotes Object of  domain $\{d_n\}$, and $\{F_{d_i}()\}$ is a function or an operation of the environment $\{d_n\}$ 

\begin{definition}
  A Role is a set of users group or subjects that can execute some given operation F() within a given domain. Each users group can be scaled down to a single user. The role $ \{R_i\} $ within a system is defined as follows:
  \[
  \{R_i\}  =  \langle U_i, F_i (), d_i \rangle 
  \]
\[
  \text{Where, $ U_i=\{\{u_j\}_j\} \mid u_j$ can execute $\{F_{d_i}(O_{d_i})\}$}
\]
\end{definition}

\begin{definition}
  Objects are the entities that can be manipulated by subjects. An object can be a record, file, data warehouse, or any entity on which an operation can be executed by a user or subject. The object $ \{O_i\} $ within a system is defined as follows:
\[
  \{O_i\}  =  \langle E_i, OP_i() \rangle
\]
\[
 (e_j\in\{E_i\}) \wedge (op_j\in OP_i() ) \Rightarrow op_j(e_j) \neq\Phi
\]

Whereas, 
   $\{E_i\}$ is a set of entities (records, files, data warehouses) and 
   $\{Op_i\}$ is a set of operations (read, write, ...)
\end{definition}

\begin{definition}
  A Conflict-relation is a relation that exist between the competing roles. Roles have conflict with each other on the bases of access granted or denied permissions on objects of different domains. Conflict of interest relation is a binary relation, and within a system is defined as follows:
\[
  \text{ CIR  $ \subseteq  \{{R}^2\} \Rightarrow \{0, 1\} $ | if $ r_i \in R = 1, $ otherwise 0 }
\]
  
Moreover, let CIR $ \subseteq \{R X R\} $ must satisfies the following properties:
\begin{enumerate}
    \item CIR-1: CIR is symmetric: If role $ R_i $ is in conflict with the role $ R_j $, then $ R_j $ is certainly in conflict with $ R_i $.
    \[
       \text{ $ \forall{(r_i, r_j)} \in R^2 ; RC_i(r_i, r_j) = RC_i(r_j, r_i) $ }
   \]
    \item CIR-2: CIR is anti-reflexive: The role cannot conflict to itself.
    \[
       \text{ $ \forall{(r_i, r_i)} $ in $ R^2 ; RC_i(r_i, r_i) = 0 $ }
   \]
    \item CIR-3: CIR is anti-transitive: If role $ R_i $ is in conflict with the role $ R_j $, and $ R_j $ is in conflict with the role $ R_k $ then role $ R_i $ may or may not in conflict with $ R_k $.
    \[
       \text{ $ \forall{(r_i, r_j, r_k)} \in R ; RC_i(r_i, r_j) = 1, RC_i(r_j, r_k) = 1 $}
    \]
    \[   \text{ Then $ RC_i(r_i, r_k) = [0, 1] $ }
   \]
\end{enumerate}
\end{definition}

\begin{definition}
  A conflict-class is defined on the notion of equivalence class, that is the subset of roles which are equivalent to each other called as roles class (RC). Whereas, each role within the class must satisfy the defined properties of conflict-relation. So that we cannot find the conflicting roles within the same class. Conflict-Class within a system is defined as follows:
\[
  \text{ $ [RC_i] = \{r_i \mid (r_j, r_i) \in CIR\} $ }
\]
\[
   \text{ $ (r_i \in \{RC_i\}) \wedge (o_i \in \{RC_i\}) \Rightarrow u_i[F_{d_i}(O_{d_i})] \neq \Phi $ }
\]
\{RC\} = set of roles which are authorized to perform functions $ \{F_{d_i}()\} $ on object $ \{O_{d_i}\}. $ \\
However,
\[
  \text{ $ (r_i \notin \{RC_i\}) \wedge (o_i \notin \{RC_i\}) \Rightarrow u_i[F_{d_i}(O_{d_i})] = \Phi $ }
\]
\{CRC\} = set of competing roles classes which are not authorized to perform functions $ \{F_{d_i}()\} $ on the objects $ \{O_{d_i}\} $ of their competing class.
\end{definition}
\begin{definition}
  A Wall is build on top of the statement "Who can access what?" in a secure manner and implemented as the object and subject walls. Based on the conflict-class, we created the binary wall that decidedly separate the conflicting subjects and objects from each others. The walls within a system are defined as follows:
\[
   \text{ \{OW\} = $ \{ \forall{(r_i, r_j)} \in R^2 \mid RC_i(r_i, r_j) = 1 \} $}
\]
Set of object walls (OW) is the set of roles $ (r_i, r_j) \in R^2 $ such that $ RC_i(r_i, r_j) = 1 $ whereas, the $ \overline{RC_i(r_i, r_j)} = 0 $
\[
   \text{ \{SW\} = $ \{ \forall{(s_i \in r_i, s_j \in r_j)} \mid RC_i(s_i, s_j) = 1 \} $}
\]
Set of subject walls (SW) is the set of subjects that belong to their relative roles/ roles classes $ {(s_i \in r_i, s_j \in r_j)} $ such that $ RC_i(s_i, s_j) = 1 $ whereas, the $ \overline{RC_i(s_i, s_j)} = 0 $
\subsection{Object Set Binary Wall (obj)}
Object set denotes to the set of all objects, where each object $ obj_i $ belongs to $ RC_i $ and associated with two subsets of RC:
\[
    \text{ $ OWRC_{obj_i} \subseteq  RC_i $}
\]
\[    
     \text{ $ OWCRC_{obj_i} \subseteq CRC_i $}
\]
Simply $ OWRC_i $ is the set of roles classes, where they have an access right to the object $ obj_i $. $ OWCRC_i $ is the set of conflicting roles classes, denied to be accessed to the object $ obj_i $. Therefore, we can represent each pairwise $ \{OWRC_{obj}, OWCRC_{obj}\} $ by a binary pairwise of arrays $ \{ BinOWRC_{obj}, BinOWCRC_{obj}\} $, named it a binary object’s wall. Each array has a size of n bits (the name of roles classes). Bits j between 1 and n are initialized as following:
\[
    \text{ $ BinOWRC_{obj} $ is 1 if $ RC_i \in OWRC_{obj} $, otherwise 0.}
\]
\[
    \text{ $ BinOWCRC_{obj} $ is 1 if $ RC_i \in OWCRC_{obj} $, otherwise 0.}
\]
The once operation authorised on bits, is to change value from 0 to 1.
\subsection{Subject Set Binary Wall (sub)}
Subject set denotes to the set of all subjects, where each subject $ S_i $ associated with two subsets of roles class (RC).
\[
    \text{ $ SWG_{S_{i}} \subseteq RC_i $}
\]
\[
    \text{ $ SWD_{S_{i}} \subseteq RC_i $}
\]
Simply $ SWG_i $ is the access granted set. It is a set of roles classes have similar objects inside the subject wall of $ S_i $. $ SWD_i $ is the denied set. It is a set of roles classes denied to will be access by the subject $ S_i $. For the binary wall creation around the subjects, we can represent each pairwise $ \{SWG_s, SWD_s\} $ by a binary pairwise of arrays $ \{BinSWG_s, BinSWD_s\} $, named it a binary subject’s wall. Each array has a size of n bits (the number of roles classes). Bit j between 1 and n are initialized as following:
\[
    \text{ $ BinSWG_s $ is 1 if $ RC_i \in SWG_s $, otherwise 0}
\]
\[
    \text{ $ BinSWD_s $ is 1 if $ RC_i \in SWD_s $, otherwise 0}
\]
The once operation authorised on bits, is to change value from 0 to 1.
\end{definition}


\begin{definition}
  Access right is a set of operations (read, write, delete etc.) that a role has to perform on domain objects is represented in the form of Table \ref{table:ART}. We considered the two access rights in our system that are read and write. However, only the authorized user can perform their assigned operations on domain objects after their access verification and the walls will be updated according to the operations explained in section \ref{sec:SCP}.
\begin{table}[h!]
\caption{Access Rights}
\centering
\begin{tabular}{|c|c|c|c|c|c|c|}
\hline\hline
 & $ R_1 $ & $ R_2 $ &  & $ \cdots $ &  & $ R_i $ \\ [0.5ex]
\hline
$ Obj_1 $ & \{r, w\} & \{r\} & & $ \{\cdots\} $ & & $ \{\cdots\} $\\
\hline
$ Obj_2 $ & \{w\} & $ \{\emptyset\} $ & & & & \\
\hline
$ \vdots $ &  & & & & &\\
\hline
$ Obj_i $ & &  & & & & $ \{\cdots\} $\\ [1ex]
\hline
\end{tabular}
\label{table:ART}
\end{table}

\end{definition}
\section{Security Check Point}
\label{sec:SCP}
Security check point is based on the data access authorization and subject and object binary walls updates.
\subsection{Access Authorization}
Subject is granted to access the object, if and only if the following access authorisation condition is verified:
\[
    \text{ $ SWG_i \cap OWCRC_j  = \emptyset $ AND $ SWD_i \cap OWRC_j  = \emptyset $} 
\]

According to the binary walls of subject and object, the access condition is mapped as following:
\[
    \text{ $ BinSWG_s \land BinOWCRC_o = \{00…00\} $ AND }
\]
\[
    \text{ $ BinSWD_s \land BinOWRC_o = \{00…00\} $ }
\]
The access is denied, if the objects (data warehouse) of competing roles find inside the same wall.
      
\subsection{Wall Updating}
There is no need to update the walls if the access is denied. However, if the subject is granted to access the object the binary walls of subject and object will be updated  immediately according to the access type (reading or writing).

\subsubsection{Read from an object}
If the subject is granted to read from an object, the binary subject wall will be update as following:
\[
    \text{ $ BinSWG_s = BinOWRC_o \lor BinSWG_s $}
\]
\[
    \text{ $ BinSWD_s = BinOWCRC_o \lor BinSWD_s $ }
\]

\subsubsection{Write to an object}
If the subject is granted to write to an object, the binary object wall will be update as following:
\[
    \text{ $ BinOWRC_o = BinOWRC_o \lor BinSWG_s $}
\]
\[
    \text{ $ BinOWCRC_o = BinOWCRC_o \lor BinSWD_s $ }
\]

\section{Data Management Module}
\label{sec:DMM}
\begin{figure} [h]
\centering
\includegraphics[scale=0.6]{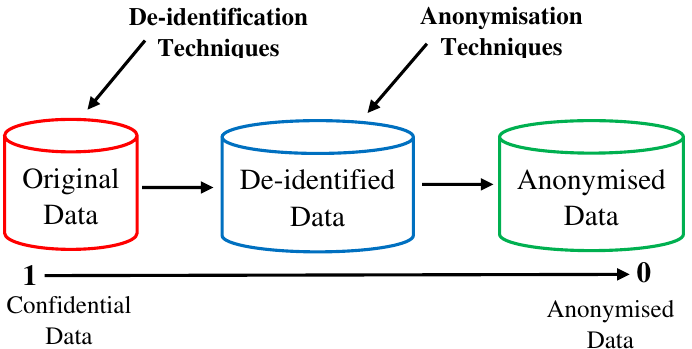}
\caption {Data Privacy Management}
\label{fig:DPM}
\end{figure}

In this section, we will present the data storage module by taking into account the privacy mechanisms as shown in Figure \ref{fig:DPM}.
\subsection{Sensitivity Level and Protection of Data}
There are four common terms that represent the nature of attributes present in the dataset on the basis of sensitivity level and relation with entity are; identifiable, Quasi, sensitive and insensitive attributes. Preserving the privacy of a piece of data that is sensitive, identifiable, or quasi-identifiable is a critical challenge. Whereas, de-identification and anonymisation of data maintains a low confidence threshold for linking the sensitive information to an individual \cite{fung2010introduction}. The most adoptable privacy models are; k-anonymity, l-diversity, (X,Y)-anonymity, LKC-privacy, t-closeness etc can be implemented using various anonymisation operations, such as noise addition, generalisation, shuffling, and perturbation\cite{sweeney2002k} \cite{machanavajjhala2007diversity} \cite{faridoon2020data}.

\subsection{Confidentiality Measurement and Data Transformation}
According to the common laws of confidentiality or data protection (Human Rights Act 1998 and Data Protection Act 1998), private data cannot be disclosed without explicit consent, legal requirement and public interest. However, these laws are based on trust, which can be easily spoiled by malicious insiders with legitimate access to confidential data. Therefore, we required data confidentiality or transformation that can prevent an individual's personal and sensitive information from someone with legitimate access to data. Data confidentiality and transformation functions within the system are defined as follows:
\[
   \text{ Data Confidentiality = $ \alpha \hspace{0.5cm} \therefore 0 <= \alpha <= 1 $}
\]
The range of $ \alpha $ from 0 to 1 measures the level of data confidentiality. This clearly indicates that $ \alpha $ = 1 is fully confidential data whereas, $ \alpha $ = 0 is less confidential and fully anonymised data.  
\[
   \text{ Data Transformation $ (F_T()) =  \{F_{D_1} \cdots F_{D_n}, F_{A_1} \cdots F_{A_m}\} $ }
\]
\[
  \text{ $ F_T(Obj_i)_{\alpha_{x}} = (Obj_j)_{\alpha_{y}} \hspace{0.5cm} \therefore \alpha_y < \alpha_x $}
\]
Data transformation $ (F_T()) $ is the set of de-identification $ (F_D) $ and anonymisation $ (F_A) $ techniques that decreases the confidentiality level and transforms the original data into anonymised data.
\subsubsection{Data De-identification}
Firstly, we consider the original data records as an object of the organization that contains the identifiable and sensitive information of an individual along with other attributes. This data is to the core confidential and must be accessible to the minimum number of roles. The Original Data within the system is defined as follows:
\[
   \text{ $ Obj_1 = OD_{{\alpha}_x} \hspace{0.5cm} \therefore \alpha_x = 1 $ (Fully Confidential Data) }
\]
Secondly, after applying the de-identification function $ (F_D) $, this original data is converted into a de-identified data object. De-identification decreases the confidentiality level of the OD.
\[
   \text{ $ F_D(OD)_{\alpha_{x}} = (DD)_{\alpha_{y}} \hspace{0.5cm} \therefore \alpha_y < \alpha_x $}
\]
\[
    \text{ $ Obj_2 = (DD)_{\alpha_y} \hspace{0.5cm} $ (Less Confidential Data) }
\]
\subsubsection{Data Anonymisation}
Thirdly, we applied anonymisation techniques on the de-identified data so that the identity of an individual cannot be revealed through linkage attacks. This data warehouse is fully anonymised and their level of confidentiality is $ \approx 0. $ 
\[
    \text{ $ F_A(DD)_{\alpha_{y}} = (AD)_{\alpha_{z}} \hspace{0.5cm} \therefore \alpha_z \approx 0 $}
\]
\[
    \text{ $ Obj_3 = (AD)_{\alpha_z} \hspace{0.5cm} $ (Fully Anonymised Data) }
\]
\section{Case Study: Model Illustration in Healthcare}
Adopting the advanced technology for Electronic Health Records (EHR) not only revolutionizes the way to treat diseases but also empowers many other sectors like; insurance companies, law enforcement agencies, pharmaceutical, and other product-selling companies etc. Based on the sensitivity, usability and multiple access of EHR, we select the healthcare sector for model illustration.
\subsection{Data Objects}
The defined data warehouses (ODW, DDW and ADW) are considered as the objects of healthcare organisation and must be protected from unauthorised access.
\subsection{Roles and Their Responsibilities}
According to the primary functions of the data science life-cycle we define certain roles and grant them access permissions regarding their responsibilities. The defined roles and their responsibilities are explained as follows (Role = R):\\
\textbf{R1: Data Collector:} Is responsible for the acquisition of patients’ data from different sources like; IoT devices, sensors, patients' health documents, physically attending patients etc. Data collection tasks can be performed by using mobile apps, web services etc there is no need to access any data warehouse.\\
\textbf{R2: Data Integration Officer:} Key concerns refer to the data quality, design and implementation of data integration applications. This will be responsible for the integration of data types and formats into a single location known as the ODW by maintaining the data quality.\\
\textbf{R3: Data Privacy Officer:} This specific role within the data management system is required for data protection. Their key responsibility is to apply data de-identification and anonymisation functions on ODW.\\
\textbf{R4: Data Controller:} involves managing the data about data (metadata), whereby this “other data” is generally referred to data models and structures, not the content. This role is ultimately accountable with regard to the definition, data quality and value of data in a given subject area.\\
\textbf{R5: Data Analyst:} By accessing the de-identified data, data analyst is responsible to structure the raw data, formulate or recognise the various patterns in the data through the mathematical and computational algorithms, and by using the standard statistical tools to analyse and interpret the data.\\
\textbf{R6: Data Scientist:} Apart from the work done by the data analyst, the data scientist will perform model creation, designing the new algorithms that can solve the specific research problems of the organisation etc. This role can collaborate with similar roles of other organisations in order to provide insights about business performance and support decision-making via sharing anonymised data.\\
\textbf{R7: End Users (Patients and Doctors etc:)} End users are not directly involved in the functions of the data management system but they can access the individual record through the interfaces defined by the healthcare organisation such as; mobile or web services, in-person services etc.
    \[
              \{U_i\} \in [R1, R4, R7] \mid U_i [F_{d_i}(ODW, DDW, ADW)] = \Phi
    \]
\subsection{Equivalent Roles Classes and Binary Walls Creation}
We assigned the roles to their particular class with respect to the access privileges they have on data objects.Table \ref{table:CIR} clearly represents the model elements.
\subsubsection{Data Management Class (DMC)}
Data Management Class contains the number of roles who can access the object (ODW) of this class. $ [DMC] = \{R2, R3\} $
\subsubsection{Data Analyst Class (DAC)}
Data Analyst Class carry the number of roles who can access the "De-identified Data Warehouse (DDW)" to perform their responsibilities. $ [DAC] = \{R5\} $
\subsubsection{Data Scientist Class (DSC)}
Data Scientist Class holds the number of roles who can fulfill their responsibilities by accessing the object ADW. $ [DSC] = \{R6\} $ \\
As their access privileges to the data warehouses, DMC, DAC, and DSC are conflicting with each others. 
\begin{table*} [t]
 \caption{Conflict of Interest Relation and Walls Representation}
 \centering
 \resizebox{\textwidth}{!}{
 \begin{tabular*}{40pc}{@{\extracolsep{\fill}}| p{2.1cm}| p{2.1cm}| p{1.1cm}| p{4.1cm}| p{1.1cm}| p{3.6cm}| @{}}
 \hline
Equivalent Roles Classes & Conflicting Roles Classes & Access Objects & Binary Object Wall \{BinOWERC, BinOWCRC\} & Subjects & Binary Subject Wall \{BinSWG, BinSWD\} \\ 
\hline
 DMC &  \{DAC, DSC\} & ODW & $ \{100, 011\} $ & S1 & $ \{100, 011\} $ \\ \hline
 DAC &  \{DMC, DSC\} & DDW & $ \{010, 101\} $ & S2 & $ \{010, 101\} $ \\ \hline
 DSC &  \{DMC, DAC\} & ADW & $ \{001, 110\} $ & S3 & $ \{001, 110\} $ \\ \hline
\end{tabular*}
 }
\label{table:CIR}
\end{table*}
\subsection{Illustration Queries}
We have following sequence queries to validate our system. All the queries are performed on binary walls.\\

\textbf{Q1}: Subject S1 needs to perform writing operation on the object ODW (original data warehouse).\\
S1 can access their required object after verifying the access authorisation condition.
\[
   \text{ $ \{100\} \land \{011\} = \{000\} $ AND $ \{011\} \land \{100\} = \{000\} $ }
\]
After the access granted to S1 and the write operation is performed on ODW, object walls will be updated as:
\[
   \text{ $ BinOWERC_{ODW} = \{100\} \lor \{100\} = \{100\} $ }
\]
\[   \text{ $ BinOWCRC_{ODW} = \{011\} \lor \{011\} = \{011\} $ }
\]
\textbf{Q2:} Subject S3 wants to perform reading operation on the object DDW. Their access is denied, because the conflicting subjects cannot access the conflicting objects. However, there is no need to update the walls because of access denied.
\[
   \text{ $ \{001\} \land \{101\} \neq \{000\} $ AND $ \{110\} \land \{010\} \neq \{000\} $ }
\]
\section{Conclusion}
As the today's world is knowledge economy, where the more you know is proportional to more data that you have. In a typical data owning organizational structure, there are number of roles/actors to whom the individuals information at each data architectural layer is accessible and the risks associated with these insiders are far more difficult because of their legitimate access to the system. Hence, we need to formulate a data privacy ensuring system to protect the identity and sensitivity of data not only from external but also from internal malicious intruders. This paper first regulated the actors/users of the organisation by adopting the strategy of RBAC. Secondly, the concept of CWSP is implemented as a firewall mechanisms to restrict the unauthorised access of users. Lastly, we proposed a comprehensive data management architecture composed of necessary components of data de-identification, and anonymisation. Therefore, our model is considered as useful and have significance in the sense that it provides security and privacy along each functional layer of data architecture. Furthermore, this system is also not subjected to faulty human intentions because it is an automatic system that impose limits on the administrators' roles with respect to regulating the accessibility of data and exercising their authority in a malicious manner. Along with this issue, it also encounters with the risk of mistakes by administrators.
\section{Acknowledgment}
This work is supported by Science Foundation Ireland under grant number $ SFI/12/RC/2289_P2 - $ Insight Centre for Data Analytics.
\bibliographystyle{IEEEtran}
\bibliography{CSCI-2022}

\end{document}